\documentclass[10pt,english]{article}
\usepackage{graphicx}
\begin{document}
\title{Remarks on the conservation of mechanical energy in introductory
mechanics}
\author{F C Santos
\footnote{e-mail: filadelf@if.ufrj.br}, V Soares \footnote{e-mail: vsoares@if.ufrj.br} and A C Tort
\footnote{e-mail: tort@if.ufrj.br.}\\
Instituto de F\'{\i}sica
\\
Universidade Federal do Rio de Janeiro\\
Caixa Postal 68.528; CEP 21941-972 Rio de Janeiro, Brazil}
\maketitle
\begin{abstract}
We discuss the work-kinetic energy theorem and the mechanical energy
conservation theorem in the context of general physics courses. The
motivation is the fact that all modern texts on introductory
mechanics show the same conceptually dangerous statement that
besides obliterating the concept of mechanical work, diminishing the
importance of the work-kinetic energy theorem, leads to erroneous
applications of the energy conservation theorem and at the same
eliminates its relationship with the principle of the conservation
of the mechanical momentum.
\bigskip
\vfill
\end{abstract}
\section{Introduction: The work-kinetic energy theorem}
Newton's laws of motion for a material point are at the very core of
classical mechanics. Nevertheless, the application of those laws to
more complex systems is not possible without the introduction of
more encompassing concepts and general theorems. Here we will
consider two of those theorems, namely, the work-kinetic energy
theorem and the mechanical energy conservation theorem. The main
motivation for this choice is the fact that almost all modern texts,
see for example \cite{Tipler1999, H&R, Feynmannlecturesvol1}, on
introductory mechanics show the same conceptually dangerous
statement that besides obliterating the concept of mechanical work,
diminishing the importance of the work-kinetic energy theorem, leads
to erroneous applications of the energy conservation theorem and at
the same eliminates its relationship with the principle of the
conservation of the mechanical momentum.
%
%
A system of particles is a set of arbitrarily chosen
particles. It is important to stress that, once the system
is chosen, no particle gets in or out of it, at least
during the time interval, finite or infinitesimally small
during which we observe it and apply to it the theorems of
classical mechanics. It is convenient to divide the particles
into two sets, the set of those that belong to the system
under study and the set of those that do not belong to the
system. The former are called internal particles and the
latter external particles. The work-kinetic energy theorem
states that the variation of the kinetic energy of a system
of particles is equal to the total work performed by all
forces acting on the system, the internal ones and the
external ones. The principle of conservation of the
mechanical energy follows from this theorem, therefore it
is not a principle at all but a demonstrable corollary to
it. However, in order to prove this corollary we must
suppose that the system is isolated and that the internal
forces are derivable from a scalar potential or do not
perform mechanical work at all. It follows from this that standard problems
such as the free fall, the frictionless sliding of a block
on the surface of a wedge, the rolling without slipping of
a sphere on an inclined plane, the gravitational catapult
and many others cannot be solved with the (mis)use of the
principle (corollary) of the conservation of the mechanical
energy. The reason is very simple: the total mechanical energy of the
system in these examples is not conserved. In what comes next
we will strive to clarify this apparently bold
statement.
\section{Some simple examples}
%
%
Consider a small block of mass $m$ sliding without friction on the
surface of a wedge with a mass $M$ such that $M\gg m$. Let us
analyse the problem initially from an inertial reference system
$\cal{S}$ fixed to the wedge. The initial height of the small block
is $h$ and for the sake of simplicity we assume that the initial
velocity is zero. The usual solution largely disseminated by the
most popular textbooks on general physics \cite{Tipler1999, H&R,
Feynmannlecturesvol1} is based on the principle (corollary) of the
conservation of mechanical energy. The reasons invoked for its
application are: (i) there is no friction in the interface between
the block and the wedge, (ii) the normal force (the constraint) does
not perform work on the block. In this way, taking as a reference a
point at the base of the wedge and recalling that the total energy
is the sum of the potential energy and the kinetic one we obtain the
following equation
\begin{equation}\label{}
mgh=\frac{1}{2}m\|\mathbf{v}\|^2,
\end{equation}
where $\mathbf{v}$ is the velocity at the base of the wedge
and parallel to it. Solving for $\|\mathbf{v}\|$ we obtain
\begin{equation}\label{}
\|\mathbf{v}\|=\sqrt{2gh}.
\end{equation}
Consider now the same problem observed from a reference
system $\cal{S}^\prime$ moving with a constant velocity
such that the velocity $\mathbf{V}$ of $\cal{S}$ with
respect to $\cal{S}^\prime$ is horizontal, i.e.: parallel
to the base of the wedge. Then the initial velocity of the
small block is $\mathbf{V}$ and its final velocity is
$\mathbf{v}+\mathbf{V}$. Therefore, the total initial
energy of the block is
\begin{equation}\label{}
E_0=\frac{1}{2}m\mathbf{V}^2+mgh,
\end{equation}
and the final energy is
\begin{equation}\label{}
 E_1=\frac{1}{2}m\left(\mathbf{v}+\mathbf{V}\right)^2.
\end{equation}
Therefore, the variation of the mechanical energy of the block is
\begin{equation}\label{variation}
 \Delta E=m\mathbf{v}\cdot\mathbf{V}.
\end{equation}
The last equation shows clearly that the total energy of the block
in the reference system $\cal{S}^\prime$ is not conserved and that
there is no sense in talking about the conservation of the energy of
the block with respect to the system $\cal{S}$ or any other
reference system. This is so because for strong physical reasons
\emph{the principle must hold in any inertial reference system}.
This point is clearly stressed at the introductory level only in
Ref. \cite{BerkeleyVol1}, to the authors' knowledge. Here, the
illusion of the conservation of the energy, according to Eq.
(\ref{variation}), holds in the $\cal{S}$ reference system only. In
this particular problem we find an additional complication, to wit,
the system is not isolated since the wedge is on an horizontal
plane. If we change the block by a rigid sphere that can roll down
along the wedge without slipping we will find the same difficulties
found in the sliding of the block. The fact that the friction forces
do not perform work as textbooks usually state is no excuse for
applying the conservation of energy. In these two examples, the
impossibility of making use of the conservation of the energy
becomes crystal clear if we take into account that in another
inertial reference system the normal force and the friction force
surely will do mechanical work.


Consider now an equally simple problem that can be viewed as an
isolated system, but again the conservation of mechanical energy is
erroneously applied to a part of the system only. Consider the free
fall of a particle of mass $m$ from a height $h$ above the surface
of the Earth and calculate its velocity immediately before it hits
the ground. Most textbooks analyse the problem from the viewpoint of
an inertial reference system $\cal{S}$ fixed with respect to the
Earth and make use of the conservation of mechanical energy to
obtain the final velocity, see for example \cite{Tipler1999, H&R}.
Besides, it is argued that since the Earth has a much larger mass
than the particle it follows that its acceleration is negligible and
therefore it can be considered as good approximation to a \emph{bona
fide} inertial reference system. The solution follows imediately
from these assumptions. We write
\begin{equation}\label{}
mgh=\frac{1}{2}m\mathbf{v}^2,
\end{equation}
and solve for $\|\mathbf{v}\|$. Consider the same problem
from a reference system $\cal{S}^\prime$ whose velocity
with respect to the Earth is $\mathbf{V}$. The initial
energy is given by
\begin{equation}\label{}
    E_0=\frac{1}{2}m\mathbf{V}^2+mgh,
\end{equation}
and the final energy is
\begin{equation}\label{}
    E_1=\frac{1}{2}m\left(\mathbf{v}-\mathbf{V}\right)^2+mg\left(-\mathbf{V}\right)\cdot
    \mathbf{\hat k}^\prime\sqrt{\frac{2h}{g}},
\end{equation}
where we have taken into account the motion of surface of
the Earth with respect to $\cal{S}^\prime$ and the
invariance of the elapsed time of free fall. The variation
of mechanical energy is now given by
\begin{equation}\label{}
    \Delta E=-m\mathbf{v}\cdot\mathbf{V}-mg\mathbf{V}\cdot
    \mathbf{\hat k}\sqrt{\frac{2h}{g}},
\end{equation}
Once again conservation of the mechanical energy holds for one
special reference system, the one in which the Earth is at rest.
\section{The sliding block problem revisited}
In the examples considered above it was shown that the mechanical
energy is not conserved in an arbitrary inertial reference system if
the mechanical system under study is not isolated. But our
conclusion is based on the galilean rule for the transformation of
velocities from one inertial system to another one. Our aim now is
to show how a consistent application of the conservation laws of
energy and momentum allow us to obtain the correct results. In order
to do this we will simplify the problem and will keep only its
essential features. We start by showing in a specific example that
the conservation theorems correctly applied lead to the conclusion
that the infinite mass of a part of the mechanical system is no
excuse for disregarding energy and/or linear momentum transfer to
it.

Consider again the problem of small block of mass $m$ sliding down
the inclined surface of a wedge of mass $M$. Choose an inertial
system with respect to which the velocity of the wedge is always
horizontal. Applying the theorems of the conservation of mechanical
energy and linear momentum to the system constituted by the wedge
\emph{and} the small block we can write
\begin{equation}\label{ConE}
\frac{1}{2}\left(M+m\right)\mathbf{V}_0^2+mgh=\frac{1}{2}M\mathbf{V}^2
+\frac{1}{2}m\left(v_x^2+v_y^2\right),
\end{equation}
and
\begin{equation}\label{ConPx}
 \left(M+m\right)V_0=MV+mv_x,
\end{equation}
where $\mathbf{V}_0$ is the initial velocity of the wedge, $h$ is
the initial height of the small block and $v_x$ and $v_y$ are the
horizontal and the vertical components of its velocity at an
arbitrary point. We  have one more equation to write that reflects a
kinematical constraint: the small block must never leave the
inclined surface of the wedge, i.e.,
\begin{equation}\label{constraint}
 v_x-V=v_y\tan\,\theta,
\end{equation}
where $\theta$ is the measure of the angle formed by the inclined
surface of the wedge and the horizontal reference line. Equations
(\ref{ConE}) and (\ref{ConPx}) can be rewritten in the form
\begin{equation}\label{ConE2}
M\left(V-V_0\right)\left(V+V_0\right)-m\left(V_0-v_x\right)
\left(V_0+v_x\right)+mv_y^2-2mgh=0,
\end{equation}
and
\begin{equation}\label{ConPx2}
M\left(V-V_0\right)=m\left(V_0-v_x\right).
\end{equation}
Combining Eqs. (\ref{ConE2}) and (\ref{ConPx2}) we obtain
\begin{equation}\label{ConE3}
m\left(V_0-v_x\right)\left(V-v_x\right)+mv_y^2-2mgh=0.
\end{equation}
From Eq. (\ref{ConPx}) we have:
\begin{equation}\label{}
    V=V_0+\frac{m\left(V_0-v_x\right)}{M}.
\end{equation}
Taking this result into Eq. (\ref{constraint}) we obtain
\begin{equation}\label{}
 v_x-V_0=\frac{Mv_y\cot\,\theta}{M+m}.
\end{equation}
If we now take this result together with Eq. (\ref{constraint}) into
Eq.(\ref{ConE3}) we can solve for $v_y$ thus obtaining
\begin{equation}\label{vy}
 v_y=\sqrt{\frac{2gh}{1+\frac{M}{M+m}\cot^2\,\theta}}.
\end{equation}
Notice that when $\theta\to\pi/2$, the above formula gives
$v_y\to\sqrt{2gh}$, and when $\theta\to 0$, $v_y\to 0$ as it must
be. The determination of $v_x$ and $V$ follows easily
\begin{equation}\label{vx}
    v_x=V_0-\sqrt{\frac{2gh}{1+\frac{M+m}{M}\tan^2\,\theta}},
\end{equation}
and
\begin{equation}\label{V}
    V=V_0+\frac{m}{M}\sqrt{\frac{2gh}{1+\frac{M+m}{M}\tan^2\,\theta}}.
\end{equation}
Let us evaluate the variation of the energy and linear momentum of the wedge. The former is given by
\begin{eqnarray}\label{deltaE}
\Delta\,E &=&\frac{1}{2}M\left(V^2-V_0^2\right)\nonumber  \\
&=&
\frac{m}{2}\sqrt{\frac{2gh}{1+\frac{M+m}{M}\tan^2\,\theta}}\left(
2V_0+\sqrt{\frac{2gh}{1+\frac{M+m}{M}\tan^2\,\theta}}\right),
\end{eqnarray}
and the latter by
\begin{eqnarray}\label{deltaP}
\Delta P &=& M\left(V-V_0\right)  \nonumber\\
&=&  m\sqrt{\frac{2gh}{1+\frac{M+m}{M}\tan^2\,\theta}}.
\end{eqnarray}
Finally we can analyse the behaviour of the system when the mass of
the wedge goes to infinity. Taking the limit of $M\to\infty$ in Eqs.
(\ref{deltaE}) and (\ref{deltaP}) we obtain
\begin{equation}\label{limE}
\lim_{M\to\infty}\Delta E=mV_0\sqrt{2gh},
\end{equation}
and
\begin{equation}\label{limP}
\lim_{M\to\infty}\Delta P=m\sqrt{2gh}.
\end{equation}
This example clearly shows how the principle of conservation of mechanical
energy is misused. In the first place, Eq. (\ref{ConE}), though correct,
cannot be interpreted as conservation of energy because the system is not isolated,
the external forces being supplied by gravity and contact forces. Equation (\ref{ConE})
is a consequence of the work-kinetic energy theorem, which, in an obvious notation,
reads for the system small block $+$ wedge
\begin{equation}\label{}
 W_{\mbox{g}}+W_{\mbox{c.f.}}=mgh=\Delta K_M+\Delta K_m.
\end{equation}
In the second place, Eq. (\ref{limE}) shows that even in the infinite mass limit, or
if one prefers, when $M\gg m$,  there is energy transfer to the wedge.
\section{Another example: The frontal collision problem}
As a second example consider a block of mass $M$ at rest with respect to the inertial reference system $\cal{S}$ and also a particle of mass $m$ moving with a velocity $\mathbf{u}$ on a head-on collision course with the block. Suppose also that there is no external force acting on the parts of the system. In the end, of course, we will be interested in the particular, but important case where the mass of the block goes to infinity ($M\to\infty$). In this limiting situation, with respect to $\cal{S}$ the final velocity of the particle is $-\mathbf{u}$ and the final velocity of the block is still null. We will demonstrate that these results cannot be obtained with the application of the principle of the conservation of the energy only. Going from the inertial reference system $\cal{S}$ to $\cal{S}^{\,\prime}$ with respect to which the velocity of the particle is $\mathbf{v}$ and the velocity of the block is $\mathbf{V}$ we write
\begin{equation}\label{}
    MV+mv=MU+mu,
\end{equation}
and
\begin{equation}\label{}
    \frac{1}{2}MV^2+\frac{1}{2}mv^2=\frac{1}{2}MU^2+\frac{1}{2}mv^2,
\end{equation}
where $U$ and $u$ are the velocities of the block and of the particle respectively, after the collision. Solving for $U$ and $u$ we obtain
\begin{equation}\label{}
    U=\frac{\left(2u-V\right)m+MV}{M+m},
\end{equation}
and
\begin{equation}\label{}
    u=\frac{\left(m-M\right)v+2MV}{M+m}.
\end{equation}
The variations of the energy and linear momentum of the block are
\begin{equation}\label{}
    \Delta E_M=\frac{M}{2}\left[\left(\frac{\left(2u-V\right)m+MV}{M+m}\right)^2-\frac{1}{2}V^2\right]
\end{equation}
and
\begin{equation}\label{}
    \Delta P_M=M\left[\frac{\left(2u-V\right)m+MV}{M+m}\right].
\end{equation}
respectively. Now take the limit $M\to\infty$ in the four equations above. The results are
\begin{equation}\label{}
    U=V,
\end{equation}
\begin{equation}\label{}
    u=v+2V,
\end{equation}
and
\begin{equation}\label{}
    \Delta E_M=4mV\left(V-v\right),
\end{equation}
\begin{equation}\label{}
    \Delta P_M=2\left(v-V\right)m.
\end{equation}
Once again it is clear that the energy and the momentum of the block change even when its mass goes to infinity, or if we prefer, when $M\gg m$. However, in the special case where the initial velocity of the block is zero ($V=0$), only its linear momentum changes.
\section{Clarifying the matter}
Consider the problem from a general point of view. Let a system of $N+M$ particles each with a mass $m_i$, $i=1...N+M$ be. Suppose that the system is isolated. Suppose also that the internal forces can be classified into two sets, namely: the set of conservative forces and the set of forces that do not perform work. Denoting by $\mathbf{x}_i$ the position and by $\mathbf{v}_i$ the velocity of of particle of mass $m_i$, we write the total mechanical energy, the total linear momentum of the system, and the total angular momentum, respectively, as
\begin{equation}\label{totalenergy}
E=\sum_{i=1}^{N+M}\frac{1}{2}m_i\mathbf{v}_i^2+U\left(\mathbf{x}_i\right),
\end{equation}
where $U\left(\mathbf{x}_i\right)$ is the total internal potential energy of the system, and
\begin{equation}\label{totallinearmomentum}
    \mathbf{P}=\sum_{i=1}^{N+M}m_i\mathbf{v}_i
\end{equation}
and
\begin{equation}\label{totalangularmomentum}
    \mathbf{L}=\sum_{i=1}^{N+M}\mathbf{x}_i\times m_1\mathbf{v}_i.
\end{equation}
Divide the system into two subsystems. The subsystem $A$ formed by the $i=1...N$ particles
and the subsystem $B$ formed by the $N+1...N+M$ remaining particles. In this way the total
mechanical energy given by Eq. (\ref{totalenergy}), the total linear momentum given by Eq.
(\ref{totallinearmomentum}) and the angular momentum can decomposed in the following way
\begin{equation}\label{totalenergydecomposition}
    E=T_A+T_B+U_A+U_B+U_{AB},
\end{equation}
\begin{equation}\label{totallinearmomentumdecomposition}
    \mathbf{P}=\mathbf{P}_A+\mathbf{P}_B,
\end{equation}
\begin{equation}\label{totalangularmomentumdecomposition}
    \mathbf{L}=\mathbf{L}_A+\mathbf{L}_B,
\end{equation}
where $T_{A\,(B)}$ is kinetic energy of the system $A\,(B)$, $U_{A\,(B)}$ is the potential
energy of $A\,(B)$, and $U_{AB}$ is the interaction potential between the two subsystems.
Recall that the complete system is isolated, therefore these mechanical quantities are conserved,
that is
\begin{equation}\label{ddtdecomposedE}
\frac{d}{dt}\left(T_A+U_A+U_{AB}\right)+\frac{d}{dt}\left(T_B+U_B\right)=0,
\end{equation}
\begin{equation}\label{ddtdecomposedP}
 \frac{d\mathbf{P}_A}{dt}+\frac{d\mathbf{P}_B}{dt}=0,
\end{equation}
\begin{equation}\label{ddtdecomposedL}
 \frac{d\mathbf{L}_A}{dt}+\frac{d\mathbf{L}_B}{dt}=0 \,.
\end{equation}
Eq. (\ref{ddtdecomposedE}) can be rewritten as
\begin{equation}\label{c.of m.}   \frac{d}{dt}\left(T_A+U_A+U_{AB}\right)=-\frac{d}{dt}\left(\frac{1}{2}M_B
\mathbf{V}_B^2+T^{\,\prime}_B\right)-\frac{dU_B}{dt},
\end{equation}
where $M_B$ and $\mathbf{V}_B$ are the mass and centre of mass
velocity of the subsystem $B$, respectively, and $T^{\,\prime}_B$ is
its kinetic energy with respect to the centre of mass. In order to
consider the subsystem $B$ as an inertial reference system suppose
that the particles belonging to $B$ are rigidly linked one to the
other. This means that its internal potential energy is constant,
$U_B=\mbox{constant}$, and the kinetic energy is given by
\begin{equation}\label{}
T^{\,\prime}_B=\frac{1}{2}\mathbf{\omega}_B\cdot\mathbf{I}\cdot\mathbf{\omega}_B,
\end{equation}
where we have introduced the angular velocity vector
$\mathbf{\omega}_B$ of the subsystem $B$ and its inertia tensor.
Expanding the rhs of Eq.(\ref{ddtdecomposedE}) we obtain
\begin{equation}\label{EABdot}
\frac{d}{dt}\left(T_A+U_A+U_{AB}\right)=-\frac{d\mathbf{P}_B}{dt}\cdot\mathbf{V}_B
-\frac{d\mathbf{L}_B}{dt}\cdot\mathbf{\omega}_B.
\end{equation}
Making use of Eqs. (\ref{ddtdecomposedP}) and (\ref{ddtdecomposedL}) we obtain
\begin{equation}\label{ddtEAB2}
 \frac{dE_{AB}}{dt}=\frac{d\mathbf{P}_A}{dt}\cdot\mathbf{V}_B
+\frac{d\mathbf{L}_A}{dt}\cdot\mathbf{\omega}_B,
\end{equation}
where we have defined the mechanical energy of the subsystem $A$ with respect to the subsystem $B$ by
\begin{equation}\label{definitionEAB}
 E_{AB}:=T_A+U_A+U_{AB}.
\end{equation}
In Eq. (\ref{ddtEAB2}), the velocities $\mathbf{V}_B$ and
$\mathbf{\omega}_B$ depend both on time. let us analyse now what
happens when the total mass of the rigid subsystem $B$ goes to
infinity. In this case after solving Eqs.
(\ref{totallinearmomentumdecomposition}) and
(\ref{totalangularmomentumdecomposition}) we obtain
\begin{equation}\label{}
    \mathbf{V}_B=\frac{\mathbf{P}-\mathbf{P}_A}{M_B},
\end{equation}
and
\begin{equation}\label{}
    \mathbf{I}_B\cdot\mathbf{\omega}_B=\mathbf{L}-\mathbf{L}_A.
\end{equation}
When we take infinite mass limit the total linear momentum
$\mathbf{P}_A$ of the subsystem $A$ remains constant, therefore we
conclude that the velocity $\mathbf{V}_B$ and the angular velocity
$\mathbf{\omega}_B$ remain constant. The total energy of the
subsystem hence is not conserved. Nevertheless, in the limit
$M_B\to\infty$ and $\mathbf{\omega}_B\to 0$, for which the velocity
of $M_B$ is zero and system B becomes an inertial system we have
\begin{equation}\label{ddtEAB}
\frac{dE_{AB}}{dt}=0.
\end{equation}
Equation (\ref{ddtEAB}) cannot be considered as a conservation law
for the mechanical energy because it holds only in the inertial
reference system for which the velocity of the subsystem $B$ is
zero. One must keep always in mind that a true conservation law must
hold for all inertial reference system. Moreover, Eq.
(\ref{definitionEAB}) that defines this energy contains the term
$U_{AB}$ that describes the interaction energy of the two systems,
hence it cannot be interpreted as the total energy of the subsystem
$A$.
%
%

%
\end{document}